\documentclass[twocolumn,showpacs,preprintnumbers,amsmath,amssymb]{revtex4}
\usepackage{amsmath,amssymb,graphics,epsfig,subfigure}
\usepackage{color}

\begin{document}

\thispagestyle{empty}

\begin{center}

\title{Universal exponents of black hole phase transition at zero-temperature limit}

\date{\today}
\author{Shao-Wen Wei$^{1,2}$ \footnote{E-mail: weishw@lzu.edu.cn}, Shan-Ping Wu$^{1,2}$, Yu-Peng Zhang$^{1,2}$,
Yu-Xiao Liu$^{1,2}$ \footnote{E-mail: liuyx@lzu.edu.cn}}

\affiliation{$^{1}$Key Laboratory of Quantum Theory and Applications of MoE, Lanzhou Center for Theoretical Physics,
	Key Laboratory of Theoretical Physics of Gansu Province,
	Gansu Provincial Research Center for Basic Disciplines of Quantum Physics, Lanzhou University, Lanzhou 730000, China\\	
	$^{2}$Institute of Theoretical Physics $\&$ Research Center of Gravitation,
	School of Physical Science and Technology, Lanzhou University, Lanzhou 730000, China}

\begin{abstract}
In this work, we investigate the universal thermodynamic characteristics of black hole phase transitions at the zero-temperature limit. Our results reveal that, far below the critical point, the near zero-temperature region also exhibits universal properties. By employing the Maxwell equal area law and analyzing the coexistence curve of black hole phase transitions, we derive three universal exponents: $\alpha=1$, $\beta=2$, and $\gamma=d-3$, where $d$ represents the spacetime dimension number. Furthermore, additional studies show that these exponents remain unchanged regardless of the black hole's charge and spin. These universal exponents provide valuable insights into enhancing our understanding of black hole thermodynamic phase transitions near zero temperature and shed light on the fundamental aspects of quantum gravity.
\end{abstract}


\maketitle
\end{center}

{\it Introduction}--- Since the pioneering work of Hawking and Bekenstein, it has been widely recognized that a black hole is not solely a manifestation of intense gravitational forces but also functions as a thermodynamic system \cite{Hawking,Bekensteina,Bekensteinb}. Taking into account the semiclassical quantum effects, a black hole can radiate particles, thus exhibiting characteristics similar to black body radiation with a temperature known as the Hawking temperature \cite{Hawking}.

As a significant thermodynamic phenomenon, the renowned Hawking-Page phase transition \cite{Hawkingpage}, which characterizes the transition between a pure thermal radiation state and a stable large black hole state, serves as a valuable tool for interpreting the gauge field's confinement/deconfinement phase transition \cite{Witten2} within the framework of the AdS/CFT correspondence \cite{Maldacena,Gubser,Witten}. Surprisingly, further investigations have uncovered stable small-large black hole phase transitions \cite{Chamblin,Kubiznak} that mirror the liquid-gas phase transition of the van der Waals (VdW) fluids. These phase transitions exhibit the similar critical  and super-critical behaviors, closely aligning with the mean field theory \cite{Kubiznak,ZhaoZhang,XuMann,LiZhang,WangLi}. Such phase transitions are universal in charged and rotating AdS black hole systems in general relativity \cite{Gunasekaran}, and they also extend to other modified gravitational models.

While the critical phenomena associated with VdW-like black hole phase transitions have been extensively explored, their prospective properties at the zero-temperature limit have yet to be fully examined. This investigation aims to unveil distinctive thermodynamic characteristics, shed light on the near-horizon geometry, and explore quantum corrections in the near-extremal black holes \cite{Iliesiu,Maulik,Kapec}. Importantly, this analysis provides insights into the realm of quantum gravity from the perspective of thermodynamics.

\begin{figure}
\includegraphics[width=6cm]{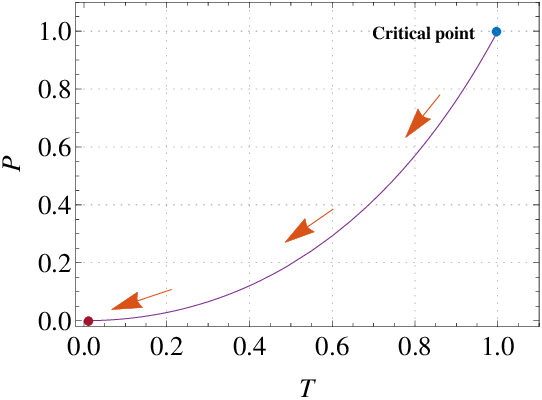}
\caption{The coexistence curve of the small-large black hole phase transition. It starts at the critical point marked with the blue dot, and extends towards low temperature and pressure, and ends at the origin marked with the red dot. In this work, we mainly focus on the universal behaviors of the thermodynamic quantities near the origin, far below the critical point.}\label{ppt}
\end{figure}

{\it Black hole phase transition.}--- To enhance our comprehension of the VdW-type black hole phase transition, we introduce the illustrative $P$-$T$ (pressure-temperature) phase diagram depicted in Fig. \ref{ppt}. The red curve represents the coexistence curve of the small and large black holes, reminiscent of the distinct phases observed in liquid and gas systems. It originates from a finite critical point indicated by a blue dot, then extends towards lower temperatures and pressures before ending at the origin denoted by a red dot. Most previous studies focus on the critical point, where critical exponents and scaling laws are revealed, and these results are consistent with mean field theory.

The VdW fluid serves as an approximate model that considers the size of molecules and their interactions. While it is effective at illustrating the liquid-gas phase transition, its efficacy diminishes in low temperatures. One key indicator of this limitation is the near-horizontal behavior of the coexistence curve for $T < 0.4T_{c}$, where $T_{c}$ represents the critical temperature. Consequently, higher-order corrections to the model become imperative. Conversely, the equations of state for black holes derive from the Hawking temperature, which incorporates semiclassical quantum effects. Thus, these equations of state are precise and could extend to lower temperature. Although modifications may be necessary for these equations in near-zero temperatures to account for complete quantum gravity, it is reasonable to expect that the equations of state for black hole systems hold the potential to unveil universal features.

Given that a system inherently attracts tends its state of lowest free energy, the coexistence of black holes can be effectively identified through the Maxwell equal area law
\begin{eqnarray}
 \oint VdP=0.\label{eqarlaw}
\end{eqnarray}
It is crucial to emphasize that $V$ is the thermodynamic volume rather than the specific volume. Considering that the volume is not a uniquely defined function of pressure, this law can be reformulated as follows:
\begin{eqnarray}
 \int_{V_1}^{V_2} PdV=P_*(V_2-V_1),
\end{eqnarray}
where $V_1$ and $V_2$ represent the thermodynamic volumes of the coexisting small and large black holes, while $P_*$ signifies the pressure at the phase transition. This transition signifies sudden changes in the extensive properties of the black holes, leading to a discontinuity in the first derivative of the system's free energy. By incorporating the equations of state, we can derive the coexistence curve. As the temperature decreases, the horizon radii $r_{hs}$ and $r_{hl}$ of the coexisting small and large black holes change inversely. To account for this, we introduce the following parameter
\begin{eqnarray}
 \chi=\frac{r_{hs}}{r_{hl}}.\label{para}
\end{eqnarray}
At the critical temperature, $r_{hs}$ equals $r_{hl}$, resulting in $\chi$ being equal to 1. Conversely, at low temperature, $r_{hs}$ is always less than $r_{hl}$ \cite{Weia}. Consequently, the parameter $\chi$ is constrained within the range of (0, 1). Particularly, in the zero-temperature limit, where $r_{hs}$ is significantly smaller than $r_{hl}$, $\chi$ tends towards 0. In the following, we will focus on these thermodynamic characteristics at small $\chi$, which corresponds precisely to the zero-temperature limit.

%%%%%%%%%%%%%%%%%%%%%%%%%%%
%%%%%%%%%%%%%%%%%%%%%%%%%%%

{\it Charged AdS black holes}---Let us consider the $d$-dimensional charged AdS black hole solutions described by the following line element
\begin{eqnarray}
 ds^2=-f(r)dt^2+\frac{1}{f(r)}dr^2+r^2d\Omega^{2}_{d-2},\\
 f(r)=1-\frac{m}{r^{d-3}}+\frac{q^2}{r^{2(d-3)}}-\frac{2\Lambda r^2}{(d-1)(d-2)}.
\end{eqnarray}
The Arnowitt-Deser-Misner mass and charge of the black holes are given by $M = (d-2) \omega_{d-2} m / 16\pi$ and $Q = \sqrt{2(d-2)(d-3)} \omega_{d-2} q / 8\pi$ with $\omega_{d-2}=2\pi^{(d-1)/2}/\Gamma((d-1)/2)$ \cite{Chamblin}. In the framework of extended phase space thermodynamics, the cosmological constant is interpreted as the pressure of the black hole, such that $P = -\Lambda / 8\pi$ \cite{Kastor}. By introducing the Euclidean imaginary time as $t = i \tilde{t}$ through a Wick rotation and expanding the line element near the horizon \cite{GibbonshHawking}, we can approximately obtain
 \begin{eqnarray}
 ds^2\simeq \rho^2d\left(\frac{2\pi \tilde{t}}{\beta_{h}}\right)^2+d\rho^2+r^{2}_{h}d\Omega^{2}_{d-2},
\end{eqnarray}
where the coordinate $\rho=\int dr/\sqrt{f(r)}$. To maintain the solution's regularity in the $\tilde{t}-\rho$ plane and eliminate the cone singularity, the imaginary time $\tilde{t}$ must have a period of $\beta_{h}$, precisely corresponding to the Hawking temperature of the black hole, given by
\begin{eqnarray}
 T&=& \frac{1}{\beta_{h}}=\frac{ 1}{4\pi}\frac{\partial f}{\partial r}\bigg|_{r=r_{h}} \nonumber\\
 &=&\frac{d-3}{4\pi r_{h}}\left(1-\frac{q^2}{r_{h}^{2(d-3)}}+\frac{16\pi P r_{h}^2}{(d-2)(d-3)}\right). \label{eom}
\end{eqnarray}
This expression is evidently closely linked to the black hole charge, pressure, and spacetime dimension. By solving it, one can derive the equation of state $P = P(T, Q, r_{h})$ for the black hole system. Alternatively, this formulation of the Hawking temperature can be interpreted as the equation of state of the black hole system. For various dimensions, the black hole systems exhibit a VdW-like phase transition. The critical point can be determined by solving $(\partial_{r_{h}}P)_{T} = (\partial_{r_{h},r_{h}}P)_{T} = 0$, around which the system displays identical critical exponents and scaling laws, in agreement with mean field theory \cite{Kubiznak}.

By utilizing the Maxwell equal area law (\ref{eqarlaw}) and the equation of state (\ref{eom}), we can successfully determine the coexisting physical quantities in terms of the parameter $\chi$. It is noteworthy that each black hole phase transition point corresponds to a unique $\chi$. Consequently, at each coexistence point, we encounter a distinct set of parameters ($P$, $T$, $r_{hs}$, $r_{hl}$) characterized by the parameter $\chi$.

\begin{figure}
\includegraphics[width=6cm]{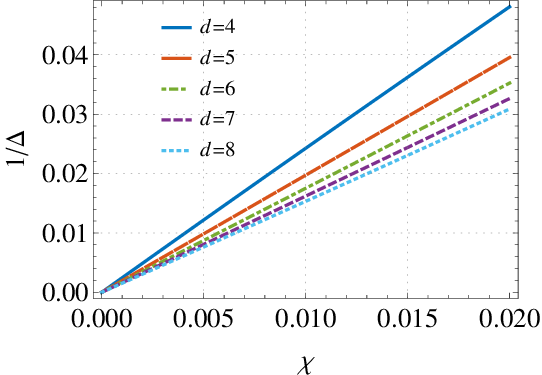}
\caption{Behavior of $1/\Delta$ as a function of $\chi$ for $d$=4, 5, 6, 7, 8 at the zero-temperature limit.}\label{pDelta10b}
\end{figure}

Let us now investigate the universal exponents at the zero-temperature limit. Regarding the black hole phase transition, the order parameter
\begin{eqnarray}
 \Delta=\frac{r_{hl}-r_{hs}}{r_{hc}},
\end{eqnarray}
exhibits a critical exponent of 1/2 in the vicinity of the critical point, where $r_{hc}$ represents the critical value of the black hole horizon radius. In this context, it is essential to investigate whether there exist universal characteristics at the zero-temperature limit. To this end, we illustrate $1/\Delta$ in Fig. \ref{pDelta10b} for $d=4, 5, 6, 7, 8$ in term of $\chi$ in low temperature. Notably, a linear relationship is evident across these dimensions, converging to $\Delta \propto 1/\chi$, irrespective of the spacetime dimension $d$. The influence of the charge $Q$ is implicit in $r_{hc}$. Specifically, we can perform an expansion of $\Delta$ around small $\chi$ at the zero-temperature limit. While the derivation is intricate, we present it explicitly for $d=4$ and $d=8$
\begin{eqnarray}
 \Delta_{d=4}&=&\frac{1}{\sqrt{6}\chi}+\frac{1}{\sqrt{6}}-\frac{7}{2\sqrt{6}}\chi+\mathcal{O}(\chi^2),\\
 \Delta_{d=8}&=&\frac{1}{66^{1/10}}\frac{1}{\chi}
 -\frac{19}{25\times 66^{1/10}}\nonumber\\
 &&-\frac{79}{1250\times 66^{1/10}}\chi+\mathcal{O}(\chi^2).
\end{eqnarray}
It is obvious that the dominant term is $\chi^{-1}$. Specifically, the spacetime dimension and black hole charge only affect the coefficients of these expansions.

%%%%%%%%%%%%%%%%%%%%%%%%%%%
%%%%%%%%%%%%%%%%%%%%%%%%%%%

Now, let us examine these thermodynamic quantities at the zero-temperature limit. For convenience, we reduce the thermodynamic quantities with their critical values, denoted as $\tau=T/T_{c}$ and $p=P/P_{c}$. By employing the parameter $\chi$ as defined in (\ref{para}), we can derive expansions for the order parameter $\Delta$ and pressure $p$ concerning the black hole phase transition at the zero-temperature limit. Following a series of calculations, we unveil the subsequent universal results
\begin{eqnarray}
 \Delta&=&a_{\alpha} \tau^{-\alpha}+\mathcal{O}(\tau^{-\alpha+1}),\\
 p&=&b_{\beta}\tau^\beta+\mathcal{O}(\tau^{\beta+1}),
\end{eqnarray}
where $a_{\alpha}$ and $b_{\beta}$ depend on the spacetime dimension $d$, and their values are listed in Table \ref{tab1} for $d=4-8$. As the dimensionality $d$ increases, $a_{\alpha}$ experiences a slight decrease, whereas $b_{\beta}$ increases and converges towards unity. Notably, across various dimensions $d$, we verify the following result
\begin{eqnarray}
 \alpha=1,\quad \beta=2.
\end{eqnarray}
Hence, we reveal the first two universal exponents at the zero-temperature limit.

Furthermore, during the transition from small to large black holes, the ADM mass undergoes a sudden change. This, to some degree, signifies a variation in the black hole's size or energy, suggesting the possible existence of a universal exponent. By defining the mass variation as $\mathcal{M}=(M_{l}-M_{s})/M_c$ with $M_{s}$ and $M_{l}$ being the masses of the coexisting small and large black holes, we expand it at the zero-temperature limit
\begin{eqnarray}
 \mathcal{M}&=&c_{\gamma} \tau^{-\gamma}+\mathcal{O}(\tau^{-\gamma+1}).
\end{eqnarray}
Of particular interest, we obtain the following result
\begin{eqnarray}
 \gamma=d-3.
\end{eqnarray}
Obviously, these exponents reveal the impact of the spacetime dimensionality $d$ on the black hole mass at the zero-temperature limit. The corresponding coefficients are also summarized in Table \ref{tab1}. Unlike $a_{\alpha}$ and $b_{\beta}$, $c_{\gamma}$ shows non-monotonic dependence on dimensionality $d$: it initially increases before undergoing a slight decrease.

In summary, for the charged AdS black hole, we have identified three exponents: $\alpha=1$, $\beta=2$, and $\gamma=d-3$. Unlike from the exponents near the critical point, these exponents characterize the universal thermodynamic behavior during the black hole phase transition zt the zero-temperature limit for the first time. Extending this methodology to other thermodynamic variables may reveal additional universal exponents.

%%%%%%%%%%%%%%%%%%%%%%%%%%%%%%%%%%%%%%%%%%%%%%%%%%%%%%%%%%%%%%%%%%%%%%%%%%%%
\begin{table}[h]
\begin{center}
\begin{tabular}{ccccccc}
  \hline\hline
       $d$& 4 & 5 & 6 & 7 & 8  \\
\hline
  $a_{\alpha}$ & 3 & $\frac{15}{8}$ & $\frac{14}{9}$ & $\frac{45}{32}$ & $\frac{33}{25}$  \\
  $b_{\beta}$ & $\frac{2}{3}$ & $\frac{64}{75}$ & $\frac{45}{49}$ & $\frac{128}{135}$ & $\frac{350}{363}$\\
  $c_{\gamma}$ &  $\frac{9}{2}$ &  $\frac{1125}{224}$ & $\frac{48020}{9477}$ & $\frac{36905625}{7340032}$ & $\frac{3013425261}{605468750}$  \\\hline\hline
\end{tabular}
\caption{Expansion coefficients at the zero-temperature limit for $d=4-8$.}\label{tab1}
\end{center}
\end{table}
%%%%%%%%%%%%%%%%%%%%%%%%%%%%%%%%%%%%%%%%%%%%%%%%%%%%%%%%%%%%%%

{\it Spinning Kerr-AdS black holes}--- The aforementioned study demonstrates the universality of the exponents $\alpha$, $\beta$, and $\gamma$ in charged AdS black hole systems. A natural subsequent question is whether these exponents remain invariant when considering rotating black holes. Previous studies have demonstrated the existence of small-large black hole phase transitions in Kerr-AdS black holes. Notably, when multiple spin parameters are introduced, reentrant phase transitions and triple points emerge \cite{Altamirano,Sherkatghanad}. However, our current investigation focuses specifically on four-dimensional Kerr-AdS black holes with a single spin parameter, though extension to include additional parameters is straightforward.

By performing the Wick rotations $\tilde{t}=i t$ and $\tilde{a}=ia$, one obtains the Euclidean Kerr-AdS geometry \cite{GibbonshHawking,Solodukhin}. By further expanding the metric near the black hole horizon and imposing the periodicity condition ($\tilde{t}$, $\phi$)$\sim$($\tilde{t}+2\pi\beta_{h}$, $\phi-2\pi\Omega\beta_{h}$) (where $a$ and $\Omega$ denote the spin and angular velocity of the black hole) yields the Hawking temperature $T=1/\beta_{h}$ of the rotating black hole
\begin{eqnarray}
 T=\frac{S^2 \left(64 P^2S^2+32PS+3\right)
     -12\pi^2 J^2}{4\sqrt{\pi} S^{3/2} \sqrt{8 P S+3} \sqrt{12 \pi^2 J^2+S^2(8 P S+3)}},
\end{eqnarray}
where $J$ and $S$ represent the angular momentum and entropy of the black hole, respectively. The analytical critical point can be determined by solving $(\partial_{S}T)_{P}=(\partial_{S,S}T)_{P}=0$ \cite{Weib}. Additionally, the coexistence curve can be obtained numerically by applying the generalized Maxwell equal area law $\oint SdT=0$ \cite{Weic}. Then, the parameter $\Delta$ can be expanded in terms of $\chi$ as
\begin{eqnarray}
 \Delta&=&\frac{0.3372}{\chi}+0.9847-1.9947\chi+\mathcal{O}(\chi^2).
\end{eqnarray}
Obviously, the dominant term is $\chi^{-1}$, consistent with the behavior of charged AdS black holes. Furthermore, for small $\tau$, the order parameter $\Delta$, the pressure $p$, and the mass change $\mathcal{M}$ exhibit the following expansions
\begin{eqnarray}
 \Delta&=&\frac{2.5707}{\tau} -1.3474 -0.3371 \tau+\mathcal{O}(\tau^2),\\
 p&=&0.7187\tau^2+0.1905\tau^3+0.0526\tau^4+\mathcal{O}(\tau^5),\\
 \mathcal{M}&=&\frac{4.0591}{\tau} -3.1935 +0.0516 \tau+\mathcal{O}(\tau^2).
\end{eqnarray}
This observation confirms the findings regarding charged AdS black holes, strongly suggesting that the spin of the black hole does not affect these universal exponents.

{\it Summary}--- In this study, we focused on analyzing the thermodynamic characteristics of black hole phase transitions at the zero-temperature limit. Significantly, by taking into account
\begin{eqnarray}
 \Delta\propto  \tau^{-\alpha},\quad
 p\propto  \tau^\beta,\quad \mathcal{M}\propto \tau^{-\gamma},
\end{eqnarray}
we obtained three universal exponents
\begin{eqnarray}
 \alpha=1,\quad \beta=2, \quad \gamma=d-3.\label{exponents}
\end{eqnarray}
This property is unaffected by the black hole's charge and spin. Moreover, the first two parameters exhibit dimension independence.

For charged AdS black holes, employing the Maxwell equal area law allows us to derive the coexistence curve analytically in terms of the parameter $\chi$ for any dimension. At each phase transition point, the black hole horizon radius undergoes discontinuous jump, resulting in variations in extensive quantities at these points. Utilizing the analytical expression of the coexistence curve, we calculated the exponents of $\Delta$ and $\mathcal{M}$ near zero temperature. The results reveal that $\Delta\sim\tau^{-1}$, a behavior distinct from that near the critical point. The exponent of $\mathcal{M}$ exhibits a $d$-dependent value of $d-3$.

For rotating Kerr-AdS black holes, the coexistence curve can also be determined using the generalized Maxwell equal area law. The numerical results precisely confirm these exponents (\ref{exponents}). This also suggests that the exponents proposed in this work are unaffected by the black hole's charge and spin.

In summary, we examined the thermodynamic characteristics and introduced three universal exponents at the zero-temperature limit. When comparing the critical exponents, our proposed exponents offer valuable insights into exploring the behavior near the event horizon of extremal black holes, potentially indicating the onset of quantum gravity effects. Moreover, this universality can be further investigated in black hole solutions that incorporate more parameters.

{\emph{Acknowledgements}}---This work was supported by the National Key Research and Development Program of China (Grant No. 2021YFC2203003),  the National Natural Science Foundation of China (Grants No. 12475055, No. 12475056, No. 12247101), the 111 Project under (Grant No. B20063), and the Gansu Province Top Leading Talent Support Plane.

\end{document}